

\input jnl-0.3.tex


\font\titlefont=cmr10 scaled \magstep3
\def\bigtitle{\null\vskip 3pt plus 0.2fill \beginlinemode \doublespace
\raggedcenter \titlefont}

\gdef\journal#1, #2, #3, 1#4#5#6{
{\sl #1~}{\bf #2}, #3 (1#4#5#6)}

\def\ap{\journal Ann. Phys., }
\def\cqg{\journal Class. Quantum Grav., }
\def\grg{\journal Gen. Rel. Grav., }

\preprintno{UCSBTH-91-44}
\def \R{{\bf R}}
\def \S{{\cal S}}
\def \e{{e_\mu}^a}
\def \ow{{\omega_\mu}^{ab}}
\def \p{\phi}
\def \a{\rightarrow}

\def \t{\times}
\def \d{\partial}
\def \l{\lambda}
\def \w{\wedge}
\def \o{\omega}
\def \de{\delta}
\def \r{\rho}

\bigtitle Topology Change in General Relativity$^*$
\body\footnote{}
{* To appear in the proceedings of the Sixth Marcel Grossman Meeting
held in Kyoto, Japan, June 24-29, 1991.}

\author Gary T. Horowitz

\affil \ucsb
\centerline{gary@voodoo.bitnet}

\abstract A review is given of recent work on topology changing solutions
to the first order form of general relativity. These solutions have metrics
which are smooth everywhere, invertible almost everywhere, and have bounded
curvature.
The importance of considering degenerate
metrics is discussed, and the possibility that quantum effects can suppress
topology change is briefly examined.

\endtitlepage

\baselineskip=16pt

\subhead{\bf 1. Introduction}

It has been over thirty years since Wheeler first suggested that the topology
of space might change in a quantum theory of gravity\refto{wh}.
However, this idea has still not been
definitely established. The difficulty is simply that
topology change does not seem to be allowed in classical general
relativity, and we do not yet have a quantum theory of gravity
which is sufficiently developed to permit reliable calculations of the
amplitude for topology to change. Over the years this issue has been considered
by
a large number of people. (It has also been investigated in a variety
of spacetime
dimensions including two\refto{gm}, three\refto{wi2}, four\refto{four},
five\refto{wi82}, and higher\refto{higher}.)

What I would like to do today is reexamine the
question of whether topology
change can occur in classical (four dimensional)
general relativity. We will see that by
slightly extending our notion of what constitutes a solution to Einstein's
equation, there {\it do} exist topology changing solutions.
Furthermore, the extension we will need is strongly motivated by quantum ideas.

Let me begin by reviewing the standard arguments for why topology change
cannot occur in general relativity. First, there is the well known result of
Geroch\refto{ge}: Consider any spacetime containing
two spacelike surfaces $\S$ and $\S'$ of different topology.
(For simplicity I will assume $\S$ and $\S'$ are  compact,
so these are closed universes, but most of the results generalize to open
universes also.)  Geroch showed in 1967
that independent of any field equations, this spacetime must have either closed
timelike curves or singularities.  There is a second result which appears to
be less well known:
If one does impose Einstein's equation or even just the weak energy
condition, $T_{\mu\nu} l^\mu l^\nu > 0$ for all null $l^\mu$,
the situation is
worse. Tipler showed in 1977 that generic topology changing spacetimes are
singular\refto{ti}. Thus, even allowing closed timelike curves,
one cannot change topology
without creating singularities.

Faced with these results, people have generally followed one of two paths.
Some have given up the local energy condition and tried to make sense of the
resulting negative energy and causality violation\refto{sor}.
However, most people have
taken a second path, which is to give up on Lorentz metrics entirely and
adopt Euclidean techniques to describe topology
change\refto{hadeg,co,gs1,garstr}.
I would like to
follow a third path  which is to keep Lorentz metrics {\it and} the field
equations
and ask what kind of singularities
are required in order to change topology? The theorems show that if a metric
satisfies Einstein's equation on a topology changing manifold, then it cannot
be well defined everywhere.
We will see that the points at which the metric is ill defined
do not have to be
strong curvature singularities. The singularities
can be very mild. In fact, they can be
so mild that in some sense they are not there at all!

   To make this more precise, it is convenient
to consider the first order form of general
relativity. This can be described in terms of a collection of four (dual)
vectors called the tetrad $\e$ and a Lorentz connection $\ow$.
The spacetime metric is recovered from
the tetrad by  $g_{\mu\nu} =\e {e_\nu}^b \eta_{ab}$ where $\eta_{ab}$ is the
constant
Minkowski metric. The action is
    $$  S=\half \int e^a \w e^b \w R^{cd}\  \epsilon_{abcd} \eqno(1)$$
where $\w$ denotes the antisymmetric wedge product,
$R=d\o + \o \w \o$ is the curvature two form and
$\epsilon_{abcd}$    is the constant antisymmetric tensor.
    The field equations  take the form
    $$   e^b \w R^{cd}\  \epsilon_{abcd} = 0 \eqno(2a)$$
    $$  e^{[a} \w De^{b]} = 0 \eqno(2b) $$
    where $D= d + \o$ is the covariant curl.
Even though these equations may look unfamiliar, I want to emphasize that this
is not an exotic new theory of gravity.
When $\e$ is invertible i.e. the tetrad consists of four linearly independent
one forms, the action (1) is completely equivalent to the usual
Einstein-Hilbert
    action and the equations of motion are equivalent to the vacuum field
    equation. Eq. (2b) says that the connection is torsion free and (2a)
    says that
    the Ricci tensor vanishes. However this action also describes
a slight extension of general relativity. Notice that the inverse of the tetrad
never appears. So
    the action and  field equations
     remain well defined even
     when $\e$ is degenerate. Thus the first order theory (1) naturally
includes
     degenerate metrics.
\subhead{\bf 2. The Importance of Degenerate Metrics}

It is conventional in general relativity to restrict consideration to metrics
which are invertible everywhere. However since general relativity is presumably
just the low energy limit of a quantum theory of gravity one should be
able to derive this restriction from the quantum theory. It is difficult
to see how this will come about. Consider the functional integral
$$   \int De D\o\  e^{i S(e, \o)}  \eqno(3)$$
One way in which degenerate metrics might be excluded is if they are infinitely
far away from nondegenerate metrics in field space. However this is not the
case\refto{giddings}.  The natural measure on the space of tetrads
is obtained from the norm
$$  ||\de e||^2 = \int \de\e \de {e_\nu}^b \ {e^\mu}_c {e^\nu}_d\  \eta_{ab}
  \eta^{cd}\ \det e
\ d^4 x  \eqno(4)$$
Even though this norm involves the inverse tetrad ${e^\mu}_c$,
the volume element more than
compensates for it. The norm (squared) is homogeneous of degree two in the
tetrad, and thus vanishes as $\e \a 0$ ensuring that $\e=0$ is a finite
distance from invertible $\e$. If only one tetrad vector vanishes, the norm
may diverge, but
the distance is still finite. For example, consider
$e^1_\mu(s) = s  e_\mu^1(0)$
and $e^a_\mu(s) = e^a_\mu(0)$ for $a \ne 1$.
Then the norm squared is proportional to $s^{-1}$,  the norm is proportional
to $s^{-\half}$, and the distance to $s=0$ is finite.

Another way to exclude degenerate metrics is to simply restrict the functional
integral (3) to invertible tetrads. However not only is this unjustified and
unnatural, it is not even gauge invariant! There are gauge transformations
which relate degenerate and nondegenerate metrics\footnote*{I wish to thank
A. Ashtekar and J. Hartle for discussions on this point.}.
To see this, we begin
with the change in the tetrad and connection under a diffeomorphism
generated by a vector field $\xi^\mu$:
$$ \delta\e =D_\mu \r^a +  2\xi^\nu D_{[\nu} e_{\mu]}^a - {\tau^a}_b {e_\mu}^b
   \eqno(5a)$$
$$ \delta \ow = \xi^\nu {R_{\nu\mu}}^{ab} + D_\mu \tau^{ab} \eqno(5b)$$
where
$$         \r^a \equiv \xi^\mu \e  \eqno(6a)$$
$$         \tau^{ab} \equiv \xi^\mu \ow \eqno(6b)$$
Since the terms involving $\tau^{ab}$ are a standard tetrad rotation, we can
consider a  modified transformation consisting of a diffeomorphism and a
compensating rotation. This simply removes the terms involving $\tau^{ab}$
from eqs. (5).
Now consider the space of solutions to the field equations (2) with
invertible metrics.
Since one of the field equations implies that the torsion vanishes
$D_{[\mu} e_{\nu]}^a =0$,  the modified transformation becomes
$$\delta\e =D_\mu \r^a  \eqno(7a)$$
$$\delta \ow = \r^c (e_c^\nu {R_{\nu\mu}}^{ab}) \eqno(7b)$$
One often considers the gauge parameter
to be a vector field $\xi^\mu$ which is independent of $\e$ and $\ow$.
($\r^a$ is then a derived quantity.) In this case it is indeed true that
finite gauge transformations do not relate degenerate and nondegenerate
metrics.
But the gauge group is actually much larger than this. The action is clearly
invariant under (5) in the more general case when $\xi^\mu$ does depend on
$\e$ and $\ow$. These are field dependent diffeomorphisms\refto{wald}.
One can thus adopt $\r^a$ as the gauge parameter. Now finite
gauge transformations do relate degenerate and nondegenerate
metrics.
For example, if we start with a flat solution ${R_{\mu\nu}}^{ab} =0$,
then the transformation (7) reduces to  $\delta \ow = 0$,
$\delta \e = D_\mu \rho^a$.
Choosing $\r^a$ to be independent of $\e$, this infinitesimal
transformation is immediately integrated to yield
$\e (s) = \e (0) + sD_\mu \rho^a$. For many choices of $\rho^a$, $\e (s)$
is degenerate for some finite
$s$\footnote{**}{Expressed in terms of $\xi^\mu$, the
vector field becomes large as the metric becomes degenerate. But this does
not seem to be sufficient justification
to rule it out. The important point is that the
change in the fields (7) remains finite.}.
More generally, we will  show in the next section
that {\it every nondegenerate solution is gauge
equivalent to one which is degenerate somewhere.}

Having hopefully convinced you of the need to consider degenerate
metrics\refto{deg}
I can now state the main result\refto{me}:
 {\it there exist smooth ($C^\infty$)
solutions to general relativity in first order form, that describe topology
change.} The tetrad $\e$
 becomes degenerate on a set of measure
 zero but the curvature remains bounded.
 (Since these topology changing solutions
 typically live on manifolds that do not admit any globally invertible Lorentz
 metric, they are not gauge equivalent to nondegenerate solutions. This shows
 that the converse of the above statement is false.)
 Thus by simply extending
 general relativity to allow degenerate metrics, one finds that topology
 change can occur classically.
 It is important to note that unlike the earlier
 Euclidean quantum gravity discussions in which topology change was viewed
 as a quantum tunneling phenomenon, the picture we obtain is of topology
 change as an essentially classical process.

 I should perhaps clarify what I mean by degenerate metrics. It is well
 known that even for the flat Minkowski metric, one can choose coordinates
 such that the determinant of the metric vanishes at certain points. This
 is a result of the fact that the new coordinates are going bad at these
points. They
 cannot be related to the original good coordinates by a smooth invertible
 transformation. This is not the sort of degeneracy which allows topology to
 change. Instead, one must fix a (topology changing) manifold with its
 collection of
 good coordinates first.  A  solution to Einstein's equation on this
 space is described by a Lorentz connection together with  four dual vectors
 which are not linearly independent at certain points. This is independent
 of the
 particular choice of (good) coordinates one chooses to express them in.
\subhead{\bf 3. Topology Changing Solutions}

To obtain the topology changing solutions we proceed as follows.
Let $M$
be a manifold with topologically different boundaries $\S$ and $\S'$.
Then one solution to the field equations
is clearly $\ow =0$, $\e =0$ everywhere.
To obtain a metric which is invertible almost everywhere, notice that
with $\ow=0$, the field equations  (2)
are satisfied if $\d_{[\mu}{e_{\nu]}}^a=0$.
So one can take any four smooth functions
$\l^a$ on $M$ and set $\e=\d_\mu\l^a$. This clearly solves the field
equations and for generic choices of $\l^a$,
the metric will be invertible almost everywhere.
To ensure that the induced metric on the boundaries is spacelike (again
almost everywhere) one need only choose $\l^0$ to be constant on the
boundaries.
The resulting metric is flat. This can be seen either from the
vanishing of the curvature two form, or by noticing that at points where
$\e$ is invertible, the metric
$g_{\mu\nu} = \d_\mu \l^a \d_\nu \l^b \eta_{ab}$ can be interpreted
as a coordinate transformation
of the constant flat metric.

One can also obtain nonflat solutions to the first order equations (2)
on any topology changing manifold. To see this, let us
first reformulate the method used to find the above solution more
geometrically. Recall that given a smooth map from one manifold $M$ to another
$\tilde M$,
one can ``pull back" $C^\infty$ covariant tensor fields on $\tilde M$ and
obtain $C^\infty$ fields on $M$. To be explicit,
if $x^\mu, \tilde x^\mu$  are local coordinates
on $M, \tilde M$ respectively, and $\tilde T_\mu $ is a $C^\infty$ one form
on $\tilde M$, then under a smooth map $\tilde x^\mu (x^\alpha)$
its pull back to $M$ is
$$    T_\alpha  = \tilde T_\mu {\d \tilde x^\mu
      \over \d x^\alpha}  \eqno(8)
	    $$
Note that this is well defined even if $\tilde x^\mu (x^\alpha)$
is not invertible.
The four functions $\l^a$ in the above solution define a smooth map from
$M$ to $\R^4$.
The solution itself is simply the pull back via this smooth
map of the standard flat Minkowski space solution on $\R^4$. Now consider any
smooth, nonsingular, curved
solution to Einstein's equation on $\R^4$. Let $\e$ and $\ow$ be the
associated tetrad
and Lorentz connection. Pick any smooth map $\p : M \a \R^4$. Then the
pull back of these forms via the
smooth map will again yield a smooth solution on $M$.
Since the map from M to $\R^4$ is not a diffeomorphism, the pull back of $\e$
can fail to be invertible even though the original $\e$ was. There will be
points where some of the one forms will vanish, and other points
where all four one forms will be nonzero but linearly dependent.
Generically, the metric will be degenerate only on a set of measure
zero.

 How does the curvature behave near these degenerate points?
 This is where one might expect this approach to break down.
 Since $\ow$ is smooth, the curvature two form is smooth everywhere. However
 curvature scalars require the inverse of $\e$ and since $\e$
 is becoming degenerate,
 one might expect these scalars to diverge as one approaches the degenerate
 points.
 But this is not the case.
 The appropriate components of the curvature two form go to zero at precisely
 the points where the tetrad becomes degenerate so that the curvature scalars
 are all finite. To see this, note that
 at all points where $\e$ is invertible, the
 curvature scalars on $M$ are just the pull back of the corresponding
 scalars on $\R^4$. Since the solution on $\R^4$ is chosen to be nonsingular,
 the curvature scalars are all finite.
 So the curvature on $M$ does not diverge as one approaches the degenerate
 points.

Now consider a solution on $M$ which is obtained via pull back under a smooth
map
$\p$ from an invertible
solution on $\R^4$. Suppose we keep the solution on $\R^4$ fixed
but change  $\p$ slightly. Then the solution on $M$ will change.
A straightforward calculation shows that the change in the solution is
precisely of the form of the gauge transformation (7) (together with a tetrad
rotation). The expression
$e_c^\nu {R_{\nu\mu}}^{ab} $ remains
well defined even when the metric degenerates,
since it is the pull back of the
corresponding expression on $\R^4$. The gauge parameter $\r^a$ is obtained as
follows. In local coordinates, a one parameter family of smooth maps is
given by a $C^\infty$ function $\tilde x^\mu(x^\alpha, s)$. If we fix
the point $x^\alpha$ of $M$ and take the derivative with respect to $s$,
we obtain a vector $v^\mu$ on $\R^4$ at the point  $\tilde x^\mu(x^\alpha, s)$.
As $x^\alpha$ changes, $v^\mu$ changes smoothly.
The gauge
parameter is $\r^a = v^\mu\e  $ where $\e$ is the tetrad on $\R^4$.
($v^\mu$ is {\it not} a vector field
on $\R^4$ since it can take more than one value at the same point, but
$\r^a$ is single valued on  $M$.)

It is now easy to prove the claim made earlier that every nondegenerate
solution to (2) is gauge equivalent to a degenerate solution under the gauge
transformations (7).
Given an invertible solution on
$\R^4$, consider a one parameter family of smooth maps $\p(s)$ from $\R^4$
to itself such that $\p(0)$ is a diffeomorphism and $\p(1)$ is
not\footnote*{This argument generalizes easily from solutions on $\R^4$
to other manifolds.}. Then the
pull back of the solution via these maps yields  a one parameter family of
solutions which interpolate between a nondegenerate and degenerate metric.
By the above remark, these are all gauge equivalent. It is important to keep
in mind that although the topology changing solutions are also obtained
by pull backs and are degenerate,
they are qualitatively different: As we remarked earlier, since $M$ typically
does not admit any globally invertible Lorentz metric, the topology changing
solutions are not gauge equivalent to
nondegenerate solutions.

We have not yet considered boundary conditions. For all topology
changing solutions we have
discussed so far it turns out that
the metric is degenerate  somewhere on each boundary.
(This is because the functions $\l^a$
must achieve a maximum and minimum on each boundary, and at these points
$\d_\mu \l^a = 0$.) It is natural to ask whether there are topology
changing solutions where the metric is invertible near each boundary. It
might appear that the answer is no: For invertible metrics, the evolution
equations of the first order theory are equivalent to the usual ones of
general relativity, which guarantee a unique invertible evolution. So if one
starts with invertible initial
data, it might appear that the solution cannot become degenerate and
the topology of space cannot change. This argument is wrong.
There do exist topology changing solutions
with invertible metrics near each boundary. To obtain them,
we will use the same pull back construction. However, rather than
mapping the boundary  into $\R^3$, we will map it into a space which
is topologically similar to it. This turns out to be a bit more delicate
than the above case. To see the potential problem,
we start with a two dimensional example.

Suppose one wants to find a
Lorentz metric on the space describing one circle spitting into two
which is flat everywhere and invertible near the boundaries (see fig. 1). One
might try to pull back the flat metric on the cylinder via a smooth map
that is the identity map on each boundary component
(so the metric will be invertible).
Unfortunately, no such map exists. There is a simple topological obstruction
which is the conservation of winding number. If $\p$ has winding number
one around each circle in the future, it must have winding number two around
the circle in the past. Fortunately, the pull back of a metric under a map
with winding number two can still be invertible. The boundary will simply
be twice as large. So one can obtain the desired metric by taking a smooth
map which is the identity on each boundary in the future and a two fold
cover in the past and pull back the flat metric on the cylinder.

For a special choice of the map $\p$ the resulting spacetime looks like fig. 2.
This is the familiar example of
two dimensional Minkowski spacetime identified under certain
translations.
Both vectors in the dyad vanish at $P$, but they are smooth everywhere.
For more general choices of $\p$, the metric
 will be degenerate
 on a circle.

We can now extend this construction to four dimensions. Recall that given
any space $\S$ with a noncontractible loop, we can obtain a new space
$\S_n$ by, roughly speaking, unwrapping this loop $n$ times. This is called the
$n$-fold covering
space. The projection back to the original space is called the covering map.
This map is the higher dimensional analog of a map with winding number
different from one. It is locally a diffeomorphism, so if a metric on $\S$ is
invertible, its pull back to $\S_n$ will be also.
We start with a solution to Einstein's
equation on $\S \t\R$. We would like to proceed as follows. First  choose
a four dimensional manifold $M$
with boundary components $\S_n$ and $\S_m$ in the future and $\S_p$ in the
past. Then
choose a smooth map from $M$ to $\S \t\R$ which reduces to the covering
map on the boundary and pull back this solution.
The question is: Does such a map exist?
Fortunately, this question has been extensively studied by differential
topologists\refto{gi,hi,cf}.
They have shown that if $ p \ne n+m$ then no smooth map exists.
(This is the higher dimensional analog of conservation of winding number.)
However, if $p=m+n$, then there always exists  a manifold $M$ with the
required boundaries such that  the map does exist.

The net result of this construction is a class of solutions describing
a compact universe which bifurcates into two, or two compact universes
which coalesce into one. For each solution
the two components are locally identical, but globally different.
I should emphasize that this is just one construction and
probably does not exhaust the class of topology changing solutions
to the first order theory with invertible metrics near the boundary.
In particular, it is likely that there exist solutions describing
bifurcating universes which are not locally identical in the future.

Although we have been working in the context of the first order form
of general relativity, it should be clear from the construction that this
is not essential. One can certainly replace the connection $\ow$ with its
self-dual part in the action (1) and still obtain topology changing solutions.
(The resulting action yields\refto{sa} Ashtekar's new variables for canonical
quantization\refto{as}.) One can also work directly with the standard second
order form of general relativity, and pull back the covariant metric
$g_{\mu\nu}$. The result is a Lorentz metric on a topology changing manifold
$M$, which is invertible almost everywhere and satisfies Einstein's
equation (in the usual sense) where ever it is invertible.
Since $g_{\mu\nu}$ is smooth, it
seems reasonable to define it to be a solution everywhere.
Using this same approach, one can clearly include matter fields as well
(at least those described by covariant tensors).

We now consider several examples.
This construction does not yield bifurcating universes with $S^3$
topology. Since a sphere is simply connected it has no covering spaces.
However there exist nontrivial examples with toroidal boundaries.
Consider a smooth solution on $T^3 \t \R$. Since all (finite) covering
spaces of $T^3$ are again topologically $T^3$, we can now obtain
a solution on a manifold with a $T^3$ boundary in the past and two disjoint
$T^3$ boundaries in the future (see fig. 3).
Note that when the boundary
are tori, the winding number around each $S^1$ is not separately conserved.
Only the total number of times the torus covers itself   matters.
As a second example let $\S$ be the three dimensional analog of a genus two
surface. In other words it is a three-sphere with two handles   added
(each topologically $S^2 \t \R$). The covering spaces $\S_n$ are now
spheres with
$n+1$ handles. Then starting with a solution on $\S \t \R$ a large number of
topology changing solutions can be constructed. The simplest example is
shown in fig. 4, which is obtained by choosing $\p$ to be a two-fold cover
in the past and the identity map in the future. Although one
cannot obtain solutions with more than two spherical boundaries this way,
one can obtain wormhole solutions (fig. 5). One can also obtain solutions
with no
boundary in the past (fig. 6). This is because the
integers $m$ and $n$
can be negative as well as positive. (A negative integer just means that
one unwraps the space in the other direction. More precisely,
the projection map from $\S_m$ to $\S$ reverses orientation rather than
preserving it.)
So one can find a solution with boundaries $\S_m$ and $\S_{-m}$  in the future.
If there is more than one noncontractible curve in $\S$, these spaces
can be topologically different.
This spacetime might be interpreted as ``pair creation of
universes from nothing". (It should be contrasted with the unique conception
theorem
of Gibbons and Hartle in the context of invertible Euclidean
metrics\refto{gh}.)
I should emphasize that in all these cases, the
metric is Lorentzian everywhere except a set of measure zero where it is
degenerate.
\subhead{\bf 4. Why is Topology Change Suppressed?}

It is clear from these classical solutions that the problem of topology change
has been turned around. The question is not whether topology change can
occur, but rather how do we stop topology from changing? Why doesn't the
space around us suddenly split into disconnected pieces?
As a first step toward resolving this embarrassing discrepancy between theory
and
experiment, let us look more closely at
the implications of these topology changing solutions for quantum
gravity. We are used to the fact that if something occurs classically,
it must occur quantum mechanically with some probability, otherwise
there is a conflict with the correspondence principle. Unfortunately,
this argument does not apply here. The reason is that it assumes unique
evolution from initial data. When the metric becomes degenerate, this fails.
(This also explains why there exist topology changing metrics which are
invertible initially.)
To see this, recall that we have constructed solutions which change
topology from $\S_{m+n}$ to $\S_m  \cup \S_n$.
Since the integers can be negative as well as positive,
one can consider the case where $n=1-m$. Now the initial surface
is identical to $\S$ (see fig. 7). This shows that there are at least
two ways to evolve initial data on $\S$: one in which the topology does
 not change and another in which it does. I want to emphasize that both
 solutions are smooth everywhere. Since for all initial data there is a
 classical evolution without topology change, there is no obvious
 inconsistency with forbidding topology changing processes in the Lorentzian
 functional integral for quantum gravity (3).

There is a possibility that the argument for topology change can be made
much stronger.
Suppose one starts with initial data on $\S$ and tries to avoid topology
change. Then one obtains a solution on $\S\t \R$. However the singularity
theorems\refto{hp} show that generically, this solution will
be geodesically incomplete.
This is usually interpreted as evidence for unbounded curvature resulting
from cosmological singularities. But in some cases, the
geodesic incompleteness is just a sign that the  metric is becoming degenerate.
  It can still have a smooth extension in the first order formalism.

For example, consider the metric
$$ ds^2 = -dt^2 + t^2 dx^2 + dy^2 + dz^2 \eqno(9)$$
where $x,y,z$ are identified with $x+1,y+1,z+1$ respectively so that
the manifold is topologically $T^3 \t \R$.
This spacetime is flat and so trivially solves
the vacuum field equation. It can be obtained by
taking the product of a flat two torus with a Lorentzian cone,
i.e. the quotient of the interior of the light cone in two dimensional
Minkowski space by a Lorentz boost. The surface $t=-1$ is a compact
Cauchy surface with the unit normals converging everywhere. By an early
singularity theorem\refto{ha1}, the spacetime must be geodesically incomplete.
Indeed, the
metric is not invertible at
the vertex of the cone, $t=0$, and so this must be removed
in standard general relativity.  However the metric is clearly smooth for all
$-\infty<t<\infty$.
In first order form,
the spacetime is described by $e^0 = dt$, $e^1 = tdx$, $e^2 = dy$,
and $e^3 = dz$. The connection can be found by solving the equation for
the vanishing of the torsion. The only nonzero component is $\o^{01} = dx$.
This is clearly a smooth solution to the first order equations
for all $-\infty < t < \infty$.

In this example, the topology of space does not change when the metric
becomes degenerate. But
there may exist more subtle examples in which one can extend the metric
smoothly
and solve the field equation only by changing the spatial topology. If so,
one can view this as a genuine
prediction of topology change in general relativity.

One mechanism for suppressing topology change has been proposed by Anderson
and DeWitt\refto{ad}. They
considered the behavior of quantum fields in topology changing backgrounds.
In particular, they investigated a scalar field propagating in the two
dimensional
geometry of fig. 2 and
found that the topology change resulted in an infinite amount of
particle creation. Further analysis\refto{dray,sorkin} has confirmed this
conclusion.
The basic
reason behind this result is the fact that a typical
scalar field becomes discontinuous as it propagates  past the degenerate point.
Since the stress tensor  is quadratic in derivatives of the field, it picks
up a term proportional to the square of a delta function which leads to an
infinite expected energy.
Since we now have a more general class of topology changing solutions, we
can ask whether this is always the case. The answer is no\footnote*{This work
was done in collaboration with A. Steif.}.
Let us stay in two spacetime
dimensions for simplicity. Then we can construct
solutions with winding number $n$ and $m$ in the future and $n+m$ in the past.
If $m$ and $n$ are both positive, then it turns out that typical
scalar field solutions will again become
discontinuous. However, if $n=1-m$, this is no longer the case. On
these topology changing spacetimes,
{\it all smooth initial data for a scalar field has a smooth evolution
for all time!} To show this, consider  any $C^\infty$
initial data on the circle. First evolve this data
on the cylinder to obtain a smooth solution. Now pull back this solution to
obtain
a smooth solution on the topology changing spacetime. Since the map on the
initial circle is the identity, the initial data is unchanged.
So the problem
that caused infinite particle production disappears. Notice that this argument
is not time reversal invariant. Given arbitrary smooth initial data on
$S_m$ and $\S_{1-m}$ in the future, there does not always exist a smooth
evolution into the past. This indicates that topology change may be a
time asymmetric process (see also ref. [\cite{dray}])
and could be a mechanism for introducing time asymmetry
into quantum gravity.

Unfortunately, the existence of smooth evolution for all initial data does
not guarantee that quantum fields are well behaved on these topology changing
backgrounds. One problem is that the evolution is not unique\footnote*{This
is related to, but not equivalent to, the lack of unique evolution
we discussed earlier. Before we considered different spacetimes evolving
from the same initial data. Here we are fixing the spacetime and considering
different evolutions of a test field in this background.}.
This was
also a problem in considering quantum fields on the topology changing
manifold shown in fig. 2. In that case,
since the field was becoming discontinuous, it was not clear what it meant for
the field to satisfy its field equation at the degenerate point. One could
essentially add an arbitrary source at that point. Requiring that the
inner product was
conserved reduced, but did not completely eliminate, this freedom in evolution.
However, it was shown that the expected energy diverged for any
evolution which conserved the inner product. For the manifold with
$n=1-m$ (fig. 7) the situation is different. Even requiring that solutions
be smooth everywhere, there are inequivalent evolutions from given initial
data. (This is related to the fact that the metric in fig. 7 is degenerate
on the surface $\gamma$ which is not contractible to a point.)
Since the field is smooth everywhere and satisfies its field equation (in the
usual sense) almost everywhere, the
inner product will be conserved for all evolutions. Unfortunately,
for any choice of evolution, it turns out that
a complete basis of functions in the past
does not evolve into a complete basis of functions in the future. This
complicates the usual calculation of particle creation.
It is not yet clear
what the effects of this are for the quantum theory. It is currently under
investigation\refto{mealan}.

To conclude, we have shown that general relativity  in first order form has
topology changing classical solutions. These solutions
have  metrics which are smooth everywhere, invertible almost everywhere
(including a neighborhood of each boundary), and
have bounded curvature. Previous arguments for infinite particle creation
do not apply to some of these solutions. It is still possible that quantum
effects will suppress topology change. If not, and if we assume
it is undesirable
or even inconsistent to explicitly forbid topology change, we are left
with a basic question: What is the ``cosmic glue" that inhibits the universe
from splitting?
\subhead{\bf  Acknowledgements}

It is a pleasure to thank the organizers of the Sixth Marcel Grossman Meeting
for the invitation to lecture. I have benefited from discussions with
A. Ashtekar, D. Eardley, S. Giddings, J. Hartle, T. Jacobson, G. Mess,
J. Polchinski, M. Scharleman, and A. Strominger. This work was
supported in part by NSF
Grant  PHY-9008502.

\subhead{\bf Figure Captions}

Fig. 1 The construction for obtaining a flat Lorentz metric on
the space describing one circle splitting into two.

Fig. 2 The solution, for a special choice of $\phi$.

Fig. 3 The construction for obtaining a solution describing one
torus splitting into two.

Fig. 4 More complicated solutions can also be obtained.

Fig. 5 A wormhole solution.

Fig. 6 A solution describing the ``pair creation of universes".

Fig. 7 Initial data on $\S$ does not have unique evolution.
The metric is degenerate on the  surface $\gamma$.

\references
\baselineskip=16pt

\refis{wh} J. Wheeler, \ap 2, 604, 1957;
  in {\it Relativity, Groups, and Topology},
  eds. B. DeWitt and C. DeWitt (Gordon and Breach, New York, 1964).

\refis{me} G. Horowitz, \cqg 8, 587, 1991.

\refis{higher} F. Tipler, \pl B165, 67, 1985; P. Mazur, \np B294, 525, 1987.

\refis{gh} G. Gibbons and J. Hartle, \prd 42, 2458, 1990.

\refis{mealan} G. Horowitz and A. Steif, in progress.

\refis{giddings} S. Giddings, ``Spontaneous Breakdown of Diffeomorphism
Invariance", preprint UCSBTH-91-23, Phys. Lett. B to appear.

\refis{deg} For further discussion of degenerate metrics see e.g.
 A. Tseytlin, \journal J. Phys. A, 15, L105, 1982;
 E. Witten, \cmp 117, 353, 1988; \np B311, 46, 1988;
 R. Percacci, \np B353, 271, 1991; R. Floreanini, R. Percacci, and
 E. Spallucci, ``Why is the Metric Nondegenerate?", preprint SISSA 132/90/EP;
and I. Bengtsson, ``Degenerate Metrics and an Empty Black Hole", preprint
90-46 ITP Goteborg, Sweden.

\refis{ad} A. Anderson and B. DeWitt, \journal Found. Phys., 16, 91, 1986.

\refis{dray} C. Manogue, E. Copeland, and T. Dray, \journal Pramana J. Phys.,
30, 279, 1988.

\refis{wald} For a discussion of field dependent diffeomorphisms,
see J. Lee and R. Wald, \jmp
   31, 725, 1990.

\refis{wi82} E. Witten, \np B195, 481, 1982.

\refis{four} In addition to the references provided below, see  e.g.
W. Kundt, \cmp 4, 143, 1967; D. Brill, in
{\it Magic Without Magic}, ed J. Klauder (W. H. Freeman, 1972); P. Yodzis,
\cmp 26, 39, 1972; \grg 4, 299, 1973; and
R. Gowdy, \jmp 18, 1798, 1977.

\refis{garstr} D. Garfinkle and A. Strominger, \pl B256, 146, 1991.

\refis{gm} J. Polchinski, \np B324, 123, 1989;
D. Gross and A. Migdal, \prl 64, 127, 1990; \np B340, 333, 1990;
M. Douglas and S. Shenker, \np B335, 635, 1990; E. Brezin and V. Kazakov,
\pl B236, 144, 1990; S. Harris and T. Dray, \cqg 7, 149, 1990.

\refis{wi2} E. Witten, \np B323, 113, 1989; S. Carlip and S. de Alwis,
\np B337, 681, 1990; Y. Fujiwara, S. Higuchi,
A. Hosoya, T. Mishima, and M. Siino, ``Topology Changes in 2+1 Dimensional
Quantum Gravity", TIT preprint TIT/HEP-170/COSMO-9 (1991).

\refis{gs1} S. Giddings and A. Strominger, \np B306, 890, 1988;
\np B307, 854, 1988.

\refis{co} S. Coleman, \np B307, 864, 1988.

\refis{ge} R. Geroch, \jmp 8, 782, 1967.

\refis{ti} F. Tipler, \ap 108, 1, 1977.

\refis{as} A. Ashtekar, \prl 57, 2244, 1986; \prd 36, 1587, 1987;
C. Rovelli, ``Ashtekar
formulation of general relativity and loop space non-perturbative quantum
gravity: a report" University of Pittsburgh preprint (1991), to appear in
Class. Quantum Grav.;  A. Ashtekar, {\sl Lectures on Nonperturbative
Canonical Gravity}, (notes prepared in collaboration with R.S. Tate)
World Scientific, 1991.

\refis{sa} J. Samuel, \journal Pramana, 28, L429, 1987; T. Jacobson
and L. Smolin, \pl B196, 39, 1987.

\refis{sorkin} R. Sorkin, ``Consequences of spacetime topology",
to appear in the proceedings of the Third Canadian Conference on
General Relativity and Relativistic Astrophysics, Victoria (1989).

\refis{sor} R. Sorkin, \prd 33, 978, 1986.

\refis{gi} C. Gibson, {\it Singular Points of Smooth Mappings} (Pitman,
London, 1979).

\refis{hi} M. Hirsh, {\it Differential Topology} (Springer-Verlag, New York,
1976).

\refis{cf} P. Conner and E. Floyd, {\it Differentiable Periodic Maps}
(Springer-Verlag, Berlin, 1964).


\refis{hp} S. Hawking and R. Penrose, \journal Proc. Roy. Soc. Lond.,
A314, 529, 1970.

\refis{ha1} S. Hawking, \journal Proc. Roy. Soc. Lond., A300, 187, 1967.

\refis{hadeg} S. Hawking, \np B144, 349, 1978;
 \pl B195, 337, 1987 .

\endreferences
\endit
\end